# Mexico-UK Sub-millimeter Camera for AsTronomy

**Castillo-Dominguez, E[1] • Ade, P[2] • Barry, P. S[2,3] • Brien T[2] • Doyle, S[2] • Ferrusca, D[4] • Gomez-Rivera, V[4] • Hargrave, P[2] • Hornsby, A[2] • Hughes, D[4] • Mauskopf, P. D.[2,5] • Moseley, P[2] • Pascale, E[2,6] • Perez-Fajardo, A[4] • Pisano, G[2] • Rowe, S[2] • Tucker, C[2] • Velazquez, M[4]**.

[1] *CONACYT / INAOE*
[2] *Cardiff University, School of Physics and Astronomy*
[3] *University of Chicago*
[4] *INAOE*
[5] *Arizona State University*
[6] *Rome University*

**Abstract:** MUSCAT is a large format mm-wave camera scheduled for installation on the Large Millimeter Telescope Alfonso Serrano (LMT) in 2018. The MUSCAT focal plane is based on an array of horn coupled lumped-element kinetic inductance detectors optimised for coupling to the 1.1mm atmospheric window. The detectors are fed with fully baffled reflective optics to minimize stray-light contamination. This combination will enable background-limited performance at $\lambda = 1.1$ mm across the full 4 arcminute field-of-view of the LMT. The easily accessible focal plane will be cooled to 100 mK with a new closed cycle miniature dilution refrigerator that permits fully continuous operation. The MUSCAT instrument will demonstrate the science capabilities of the LMT through two relatively short science programmes to provide high resolution follow-up surveys of Galactic and extra-galactic sources previously observed with the Herschel space observatory, after the initial observing campaigns. In this paper, we will provide an overview of the overall instrument design as well as an update on progress and scheduled installation on the LMT.

## 1 Introduction

The Mexico-UK Sub-millimeter Camera for AsTronomy (MUSCAT) is one of the 2$^{nd}$ generation of continuum cameras for the Large Millimeter Telescope (LMT), the largest single-dish millimeter telescope in the world operating at 1.1 mm. It is located at the top of the Sierra Negra extinct volcano at 4600 meters. The altitude of the site provides one of the driest places for millimeter observations, presenting opacities as low as 0.06 in winter [1], as demonstrated in Fig. 1.

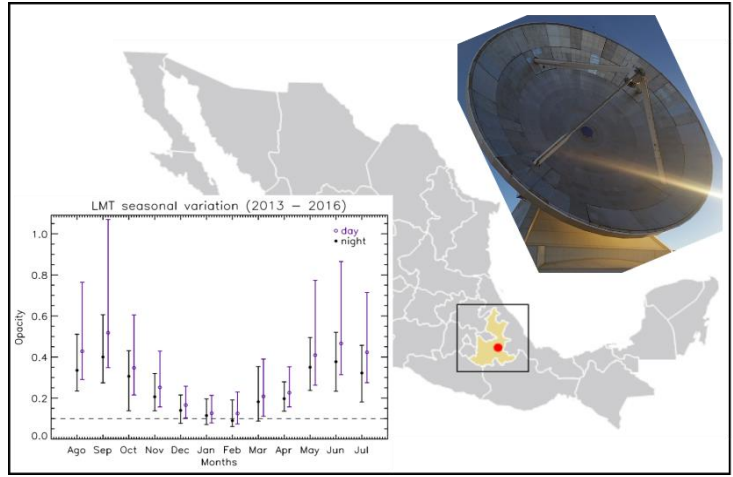

**Fig. 1** *Central* Map with the LMT site location, *Top Right* a picture of the primary surface and the tetrapod supporting the secondary mirror on the and *Bottom Left* the opacity measured along 3 years in the site [1].

MUSCAT will be a diffraction limited camera consisting of 1600 kinetic inductance detectors (KIDs) observing in the 1.1 mm atmospheric band designed to efficiently couple to both polarizations hence maximizing photometric sensitivity. Primarily, it will perform high-resolution, wide-field surveys that will probe the astrophysics of Galactic and extra-galactic sources through a dedicated observing campaign. After the primary observations are complete, MUSCAT will remain at the LMT and become an "on telescope test bed" for emerging technologies to be proven and deployed conceptually before their application in new instruments and other telescopes.

Combining the unparalleled collecting area of the LMT with the sensitivity of a highly-multiplexed KID focal plane, MUSCAT will deliver unique angular resolution and mapping capabilities leading to high-impact and important observations that will inform models of structure formation and evolution on both Galactic and extra-galactic scales.

**2 Science Case**

About 50% of the energy emitted since the Big Bang by stars and galaxies has been absorbed by dust and re-radiated at sub-mm and mm wavelenghts. The percentage is higher in the epoch when most of stars were formed at $z > 1$. This means that there is a large amount of information of the initial phase of the structure of the Universe contained in the sub-millimeter sky. The limitation in size, sensitivity and field of view of previous ground based and space borne mm wave telescopes have added biases to the results [2]. The

STFC Science Roadmap Challenges "How did the Universe begin and how is it evolving", the Astronomy Decadal Survey 2010, "New Worlds, Now Horizons" and the biases in the existing surveys, motivate the need for higher sensitivities, larger format cameras and higher resolutions.

**2.1 Observations of the Herschel-ATLAS survey**
Using the conservative predicted sensitivity of MUSCAT at 1.1 mm (table 1), enables mapping of 160 deg$^2$ of the H-ATLAS GAMA field in 40 hours [3]. As deep NIR and optical images already exist of the field, the MUSCAT survey will enable identification of the optical counterpart of all H-ATLAS sources. The redshift distribution of the H-ATLAS sources is predicted to peak at z=1, with a tail distribution to z=3 [4]. By mapping these sources with the high resolution of the LMT, MUSCAT surveys will make an investigation of the galaxy evolution over the last ~12 billion years possible.

**2.2 Observations of the Herschel-HiGal survey**
One of the revelations of Herschel is the ubiquity of interstellar filaments in star-forming regions and their close link to the formation of cores and stars. However, the study is limited by the number of the regions that can be probed with high enough resolution with current facilities. These regions are not representative due the lack of massive star formation in them.
MUSCAT installed on the LMT will perform a survey on large volumes of the Galaxy with enough spatial resolution to establish, for the first time, the global picture of the star formation process in the Milky Way.

**Table 1: Key parameter of MUSCAT on the LMT**

| Parameter | Value |
|---|---|
| Central wavelength | 1.1 mm |
| Bandwidth | 20% |
| Beam size | 5" FWHM |
| NEFD | 1.5 mJy s$^{1/2}$ |
| Number of detectors | 1600 |
| Mapping speed | 3 deg$^2$ mJy$^{-2}$ hr$^{-1}$ |

**3 MUSCAT Description**
The MUSCAT focal plane has been designed to cover the full 4 arcminute field-of-view of the LMT. It will consist of 1600 horn-coupled lumped-element KIDs (LEKIDs) [5] cooled to an operating temperature of 100 mK. The detector focal plane will be formed using a mosaic of 4 sub-arrays each with 400 KIDs. Each sub-array will be fabricated from a single 4-inch wafer.

Following the radiation path from the telescope, light gathered by the primary optics will be directed toward MUSCAT using a flat pick-off mirror (Fig. 2). Two powered warm mirrors in a crossed-Dragone configuration then form a pupil (cold stop) inside the cryostat. A strong baffling scheme was selected to minimize the effects of stray-light, as shown in Fig. 3. After the cold stop, the final image is formed at the feed horn array by two cryogenic mirrors housed inside the 350 mK superconducting shield reducing the instrument thermal emission to negligible levels and providing a magnetically stable environment to house the detectors.

To reach the 100 mK operating temperature, four cryogenic stages are required: 50 K, 4 K, 1 K and 350 mK. The first two stages will be provided by a Cryomech Pulse Tube Cooler (PTC) model 420 which has 2W of cooling power at 4 K. A 1 K continuous buffer stage is supplied by a Chase Research Cryogenics [6] and is used to pre-cool wiring and structural supports for the subsequent stages. The 350 mK stage is provided by a similar continuous dual He-10 fridge, which provides 500 μW of cooling power sufficient to cool the cold optics and radiation shields as well as providing a condensing stage for the 100mK cooler. The 100 mK stage is provided by a closed-cycle mini-dilutor fridge capable of cooling down to 88 mK with the predicted load of around 1 μW. The last three stages were ordered with Chase Research Cryogenics and will be delivered by February 2018. A detailed explanation on the cryogenic fridges is given in [7].

Arrays of LEKIDs offer numerous advantages for large-format arrays and have been successfully demonstrated at mm-wavelengths [8]. KID arrays enable multiplexing ratios of several hundreds of detectors on a single transmission line and a cryogenic amplifier, removing the need for complex readout components to be cooled to cryogenics temperatures. The readout hardware is commercially available at moderate cost and is based on ROACH2 boards powered by XILINX FPGA chips. MUSCAT will have 4 identical channels to readout each of the four sub-arrays. To date, the four ROACH2 boards along with a spare have been delivered and the integration of the readout is currently underway.

MUSCAT has been designed as a modular system in such a way to allow future upgrades to be easily implemented. In the longer term, MUSCAT will platform compatible with new technologies such as on-chip spectrometers platform compatible with new technologies such as on-chip spectrometers [9] requiring detector NEPs ~ $4\times10^{-18}$ W/Hz$^{1/2}$. Aluminum KID technologies have demonstrated NEPs ~ $10^{-19}$ W/Hz$^{1/2}$ [10] at 100mK hence MUSCAT's cryogenic capabilities are suitable for any conceivable future KID based detector to be installed and tested. MUSCATs fully reflective optics can operate across several mm / submm atmospheric windows without modification and offer diffraction limited performance out to wavelengths as short as 850 microns.

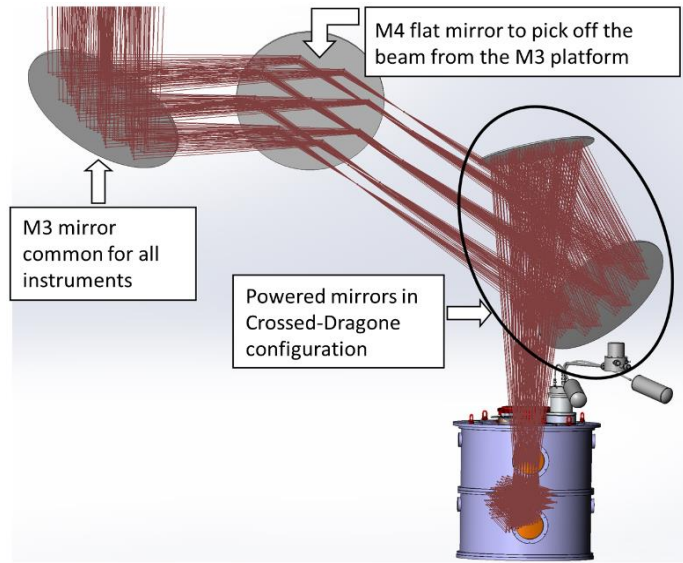

**Fig. 2:** Warm optical coupling subsystem comprising 3 large mirrors: one flat for pick off mirror and two powered mirrors to refocus the radiation into the MUSCAT cryostat.

**4 Conclusion**
MUSCAT is a KID based mm-wave camera scheduled to be installed at the LMT in Mexico in summer of 2018. The MUSCAT focal plane will cover the full telescope 4 arcminute FOV with diffraction limited performance and background limited sensitivity. It is on course to become the first instrument taking advantage of the full 50 m of the LMT. The proposed scientific programs will complement existing observations at other wavelengths and are set to provide important observations crucial to the understanding of structure formation and evolution throughout the history of the Universe.

**Acknowledgements**. We acknowledge RCUK and CONACYT through the Newton Fund (grant no. ST/P002803/1). To CONACYT for support the fellowship for the instrument scientist (grant no. 053), to Chase Research Cryogenics for the development of the fridges and to XILINX Inc. for the donation of the FPGAs used for the ROACH2 boards.

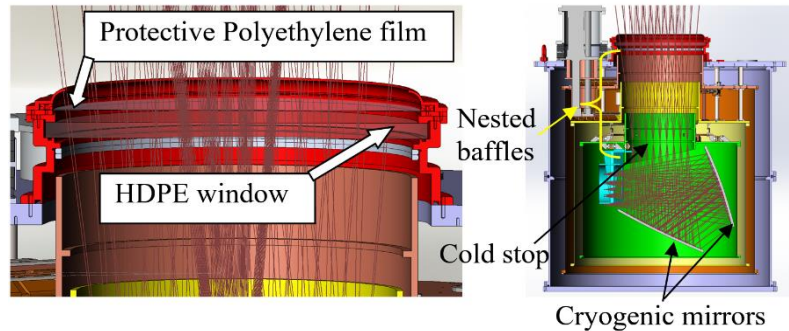

**Fig. 3:** Cross-section view of the MUSCAT cryostat highlighting the window on top, the cold stop, the aggressive nested baffling scheme at 50K, 4K and 350 mK and the cryogenic mirrors to form the image on the focal plane array.

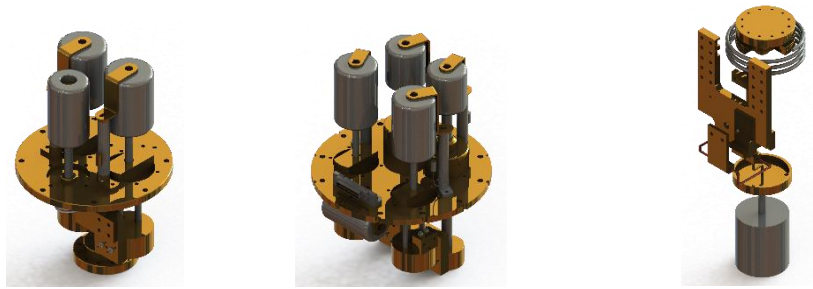

**Fig. 4:** The three closed cycle fridges to cool down from 4K to 100mK. Left to right: the 1K continuous operation fridge, the 350 mK for continuous operation and the 100 mK mini dilution fridge. All of them will be supplied from Chase Research Cryogenics, the first one has been delivered.

**References**

1. D. Ferrusca and R.J. Contreras; Proceedings of the SPIE, Volume 9147, id. 914730 10 pp. (2014). DOI 10.1117/12.2055005
2. Le Floc'h et al (2005). Infrared luminosity functions from the Chandra Deep Field-South: The Spitzer View on the History of Dusty Star Formation at 0 < z < 1. ApJ. 632(1), 169-190.
3. Eales, S. et al (2010) The Herschel ATLAS. PASP 122, 499-515.
4. Pearson et al (2013). H-ATLAS: estimating redshifts of Herschel sources from sub-mm fluxes. MNRAS, 435(4), 2753-2763.
5. Doyle, S. et al. Lumped element kinetic inductance detectors for far-infrared astronomy. Proceedings of the SPIE, Volume 7020, article id. 70200T, 10 pp. (2008).

6. Klemencic, G.M. et al. A continuous dry 300 mK cooler for THz sensing applications. Rev. of Sci. Inst. 87, 045107 (2016).
7. Brien, T. et al. A Continuous 100-mK Helium-Light Cooling System for MUSCAT on the LMT. J. Low Temp. Phys. This Special Issue (2017).
8. Calvo, M. et al., 2016. The NIKA2 Instrument, A Dual-Band Kilopixel KID Array for Millimetric Astronomy. *Journal of Low Temperature Physics*, 184(3-4), pp.816–823.
9. Wheeler, J. et al (2018), SuperSpec, The On-Chip Spectrometer: Improved NEP and Antenna Performance, *Journal of Low Temperature Physics, pp 1-7,* doi: /10.1007/s10909-018-1926-z
10. de Visser, P.J., et al., Evidence of a Nonequilibrium Distribution of Quasiparticles in the Microwave Response of a Superconducting Aluminum Resonator, Physical Review Letters, 2014. 112(4): p. 047004.